\documentclass[useAMS,usenatbib]{mn2e}
\usepackage{psfig}

\title[Optical counterpart of HLX-1]{Optical counterpart of HLX-1 during the 2010 outburst}
\author[R. Soria et al.]{{R. Soria$^{1}$\thanks{E-mail:
roberto.soria@icrar.org (RS); pahakala@utu.fi (PJH); ghau@eso.org (GKTH); 
jgladsto@ualberta.ca (JCG); akong@phys.nthu.edu.tw (AKHK).}, 
P. J. Hakala$^{2}$,
G. K. T. Hau$^{3}$,
J. C. Gladstone$^{4}$,
A. K. H. Kong$^{5}$}\\
$^{1}$International Centre for Radio Astronomy Research, Curtin University, 
GPO Box U1987, Perth, WA 6845, Australia\\
$^{2}$Finnish Centre for Astronomy with ESO (FINCA), V\"ais\"al\"antie 20, University of Turku, FIN-21500 Piikki\"o, Finland\\
$^{3}$European Southern Observatory, Alonso de C\'ordova 3107, Santiago, Chile\\
$^{4}$Department of Physics, University of Alberta, Edmonton, Alberta, T6G 2C7, Canada\\
$^{5}$Institute of Astronomy and Department of Physics,
National Tsing Hua University, Hsinchu 30013, Taiwan}
\begin{document}

\date{Accepted .... Received ...; in original form ...}

\pagerange{\pageref{firstpage}--\pageref{lastpage}} \pubyear{2011}

\maketitle

\label{firstpage}

\begin{abstract}
We studied the optical counterpart of the intermediate-mass black hole 
candidate HLX-1 in ESO\,243-49. We used a set of Very Large Telescope 
imaging observations from 2010 November, integrated by {\it Swift} X-ray 
data from the same epoch. We measured standard Vega brightnesses 
$U = 23.89 \pm 0.18$ mag, $B = 25.19 \pm 0.30$ mag, 
$V = 24.79 \pm 0.34$ mag and $R = 24.71 \pm 0.40$ mag.
Therefore, the source was $\approx 1$ mag fainter in each band than 
in a set of {\it Hubble Space Telescope} images taken 
a couple of months earlier, when the X-ray flux was a factor of 2 higher. 
We conclude that during the 2010 September observations, the optical 
counterpart was dominated by emission from an irradiated disk (which 
responds to the varying X-ray luminosity), rather than by a star cluster 
around the black hole (which would not change).
We modelled the Comptonized, irradiated X-ray spectrum of the disk, 
and found that the optical luminosity and colours 
in the 2010 November data are still consistent 
with emission from the irradiated disk, with a characteristic 
outer radius $r_{\rm out} \approx 2800 r_{\rm in} \sim 10^{13}$ cm 
and a reprocessing fraction $\approx 2 \times 10^{-3}$.
The optical colours are also consistent with a stellar population 
with age $\la 6$ Myr (at solar metallicity) and mass 
$\approx 10^4 M_{\odot}$; this is only an upper limit to the mass, 
if there is also a significant contribution from an irradiated disk. 
We strongly rule out the presence 
of a young super-star-cluster, which would be too bright. 
An old globular cluster might be associated with HLX-1, as long as its mass 
$\la 2 \times 10^6 M_{\odot}$ for an age of 10 Gyr, but 
it cannot significantly contribute to the observed 
very blue and variable optical/UV emission.
 
\end{abstract}


\section{Introduction}

The point-like X-ray source 2XMM\,J011028.1$-$460421 (henceforth, 
HLX-1 for simplicity) is the strongest intermediate-mass black hole 
(IMBH) candidate known to date \citep{far09,dav11,ser11}.
It is seen in the sky at a distance of $\approx 8\arcsec$ 
from the nucleus of the S0 galaxy ESO\,243-49 (redshift $z = 0.0224$, 
luminosity distance $\approx 95$ Mpc, distance modulus $\approx 34.89$ mag; 
at this distance, $1\arcsec \approx 460$ pc).
Its X-ray luminosity and spectral variability \citep{far09,god09,ser11} 
are consistent with high and low states of 
an accreting BH. It shows recurrent outbursts every $\approx 370$ d 
(seen in 2009 August, 2010 August, 2011 August), due either to some kind 
of disk instability, or to a periodic enhancement 
of the accretion rate.

HLX-1 has an optical counterpart \citep{sor10}, with H$\alpha$ emission 
at a redshift consistent with that of ESO\,243-49 \citep{wie10}, 
which strongly suggests a true physical association. Henceforth, 
we will assume that HLX-1 and ESO\,243-49 are located at the same luminosity 
distance. The peak X-ray luminosity of HLX1 is $\approx 10^{42}$ erg s$^{-1}$, 
three orders of magnitude higher than typical stellar-mass X-ray binaries,  
and one order of magnitude higher than the most luminous 
ultraluminous X-ray sources (ULXs) found to-date. 
The BH mass required to be consistent 
with the Eddington limit is $\sim 10^4 M_{\odot}$, and a similar value 
is obtained from spectral modelling of the thermal X-ray component, 
which is consistent with emission from an accretion disk 
\citep{far09,dav11,ser11}.

One of the most intriguing unsolved questions about HLX-1 is where it came 
from. There are no other luminous X-ray sources in ESO\,243-49, in 
the range $\sim 10^{39}$--$10^{41}$ erg s$^{-1}$; in fact, S0 galaxies 
do not generally contain extreme ULXs, which are more frequently found 
in spiral, irregular or interacting galaxies, with recent star formation 
\citep{swa04,wal11}.
For example, ULXs as luminous as $\approx 10^{41}$ erg s$^{-1}$ have been 
seen in the Cartwheel galaxy \citep{piz10}, representing the upper end 
of a rich population of luminous X-ray sources, probably high-mass 
X-ray binaries. Instead, HLX1 appears to be an odd case within ESO\,243-49.

Moreover, if the BH mass estimates are correct, HLX-1 is way too massive 
to have been formed from any stellar evolution process. Two conservative 
scenarios are that HLX-1 may be: a) the nuclear BH of a disrupted 
dwarf satellite galaxy, accreted by ESO\,243-49 \citep{kin05}; or, 
b) an IMBH formed inside a massive star cluster, perhaps via stellar 
coalescence and mergers \citep{por02,fre06}. 
The location of HLX-1 outside the disk plane of ESO\,243-49 
is consistent with both scenarios---in fact, massive globular clusters 
such as $\omega$ Cen in the Milky Way may themselves be the remnants 
of nucleated dwarf galaxies \citep{bek03}. 
IMBHs in massive star clusters, or floating 
in the outskirts of galaxies after minor mergers \citep{ole09}, 
have been predicted from theoretical arguments, 
or speculatively inferred from dynamical 
modelling \citep{geb05,lut11}, 
but their existence is far from proven \citep{van10}
and they have never been confirmed directly 
as luminous, accreting sources. 
Measuring the luminosity and colours of the optical counterpart, 
and their long-term variability, can help us test those scenarios.
The optical emission may come from: 
the outer regions of the BH accretion disk; 
a star cluster around the BH;
a single, massive and strongly irradiated donor star; 
a combination of all three. 
If the optical counterpart is dominated by a star cluster, we want 
to constrain its mass and age, which may give us a clue on the IMBH 
formation process. 

In the rest of this paper, we present 
the results of our optical photometric study with the European Southern 
Observatory (ESO)'s Very Large Telescope (VLT). We use the optical 
and X-ray luminosities and colours to model the spectral energy distribution, 
and discuss the most likely scenarios for the optical emission. 
We also compare the optical 
brightness in our VLT images with the brightness measured in the 
{\it Hubble Space Telescope (HST)} observations of \citet{far11}, taken 
when the X-ray luminosity of the source was higher.

\section{VLT observations}

We observed HLX-1 with the VIsible MultiObject Spectrograph (VIMOS) 
mounted on the Nasmyth focus B of UT3 Melipal, one of the four 
8-m VLT telescopes at Cerro Paranal. We took three images in each of the 
$UBVRI$ filters. Our observations were carried out in service mode, 
and were split between the night of 2010 November 7  
(all $U$, $V$ and $I$, and two out of three $B$ observations) 
and 2010 November 26 (the last $B$ and all $R$ observations). 
All observations were taken in photometric conditions 
with an airmass $\la 1.2$ and typical seeing $\approx 0\arcsec.7$--$0\arcsec.8$.
Standard fields were observed on each night.
See Table 1 for a more detailed log of our observations.

\begin{table}
\begin{center}
\begin{tabular}{lccr}\hline
\hline
Date &  Filter & MJD$_{\rm start}$ & Exp time\\
\hline\\[-5pt]
2010-11-07 & U & 55508.127  &  1560 s\\
   & V & 55508.147 & 375 s\\
   & V & 55508.152 & 375 s\\
   & V & 55508.157 & 375 s\\
   & U & 55508.164 & 1350 s\\
   & U & 55508.180 & 1350 s\\
   & I & 55508.198 & 165 s\\
   & I & 55508.201 & 165 s\\
   & I & 55508.203 & 165 s\\
   & B & 55508.208  &  560 s\\
   & B & 55508.215 & 560 s\\
2010-11-26   & B & 55527.139 & 560 s\\
   & R & 55527.148 & 285 s\\
   & R & 55527.152 & 285 s\\
   & R & 55527.156 & 285 s\\
\hline
\end{tabular} 
\end{center}
\caption{Log of our VLT VIMOS imaging observations.}
\label{tab1}
\end{table}

The total exposure time was 4260 s in $U$, 1680 s in $B$, 1125 s in $V$, 
$855$ s in $R$ and 495 s in $I$. In all observations, 
HLX-1 was located approximately at the same position in Quadrant 1 
of the VIMOS field. We used a standard dithering pattern for the three 
observations in each filter, to correct for pixel-to-pixel variations. 
For each filter, the three images were median-combined, and standard 
calibration procedures (cosmic ray filtering, bias subtraction, flat fielding) 
were applied through the ESO pipeline. In this paper, we focus on the Q1 
imaging data, although we also extracted and analysed the data in the other 
three VIMOS chips, which cover a significant fraction of the Abell 
cluster 2877. We used {\small {SExtractor}} \citep{ber96} 
to build a catalogue of detected objects for each of the five median-combined 
images. We then used {\small {SCAMP}} and {\small {SWarp}} 
\citep{ber02,ber06} to read the {\small {SExtractor}} catalogues, 
compute astrometric solutions, and align the images in the various filters. 

To convert from count rates to standard magnitudes (Vegamag) 
and fluxes at the top of the atmosphere, we used the standard relation
\begin{displaymath}
{\rm Mag} = -2.5 \log\left({\rm {Flux}}\left[e^{-}/{\rm {s}}\right]\right) 
    + C_1 {\rm {Col}} - C_2 {\rm {Airmass}} + {\rm ZP},
\end{displaymath}
where 
$C_1$ is the colour coefficient suitable for a particular band, 
$C_2$ is the extinction coefficient, and ZP is the zeropoint 
(different for each filter, in the Vegamag system).
We used a gain of $1.76 e^{-}/{\rm {ADU}}$ to convert from directly measured 
fluxes in ADUs/s to fluxes in $e^{-}$/s.
Zeropoints for our two nights were obtained from the pipeline-calibrated data 
files, and we also checked that they were consistent with the average values 
over the second semester of 2010, as listed on the 
ESO VIMOS quality control website
\footnote{www.eso.org/observing/dfo/quality/VIMOS/qc/zeropoints.html}.
Average colour and extinction coefficients were obtained from the same website 
(summarized in Table 2). Not enough standard fields were observed 
in the two nights to permit us to determine the extinction coefficients 
on those specific nights; however, the error introduced by taking 
average colour and extinction coefficients is negligible 
compared with other sources of error.  
Finally, we need to correct for line-of-sight Galactic extinction 
in the direction of ESO\,243-49, and any additional extinction intrinsic 
to the optical counterpart of HLX-1, in order to determine its true 
luminosity and colours. We will discuss this in Section 4. 


\begin{figure}
\begin{center}
\psfig{figure=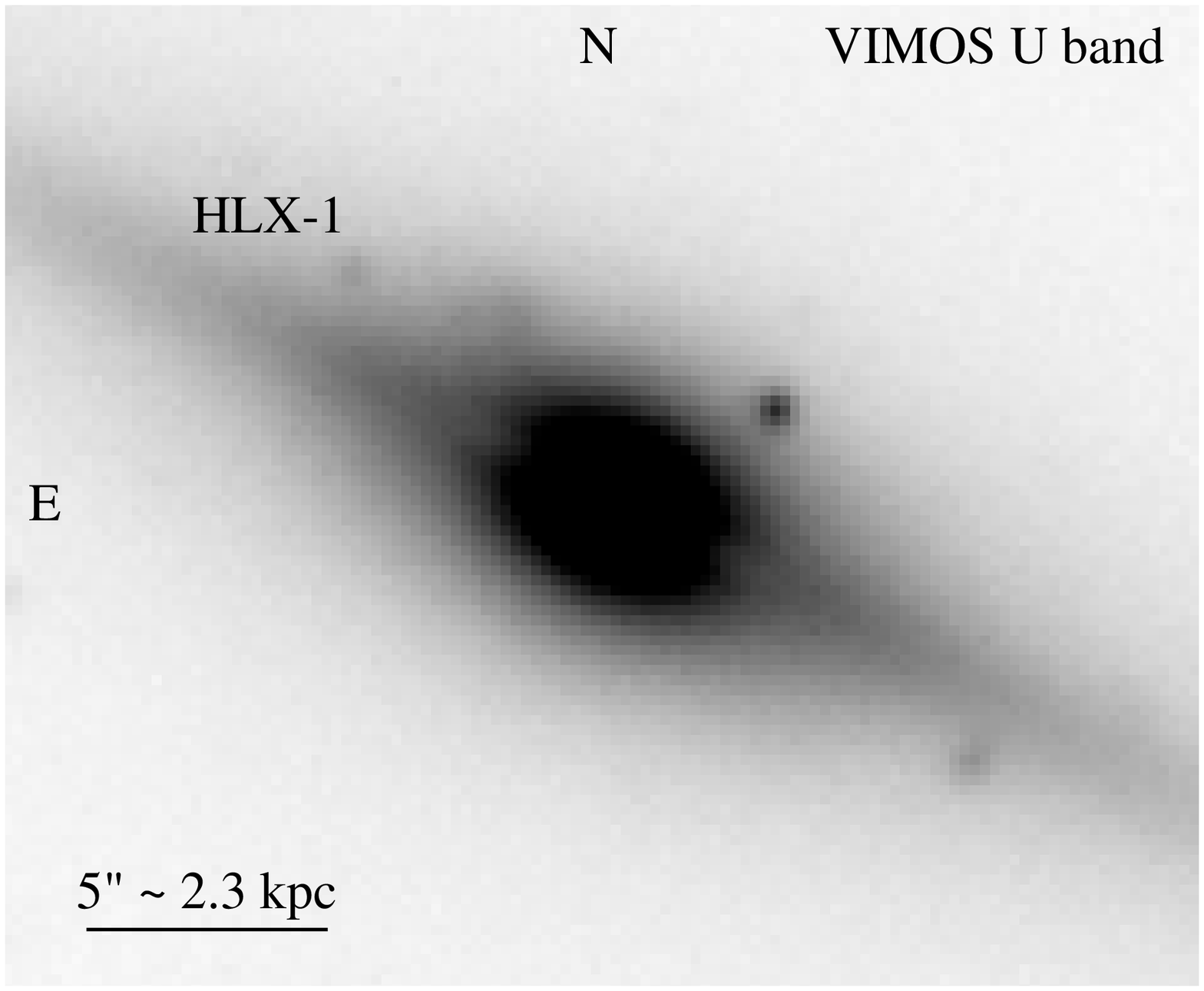,width=84mm,angle=0}
\psfig{figure=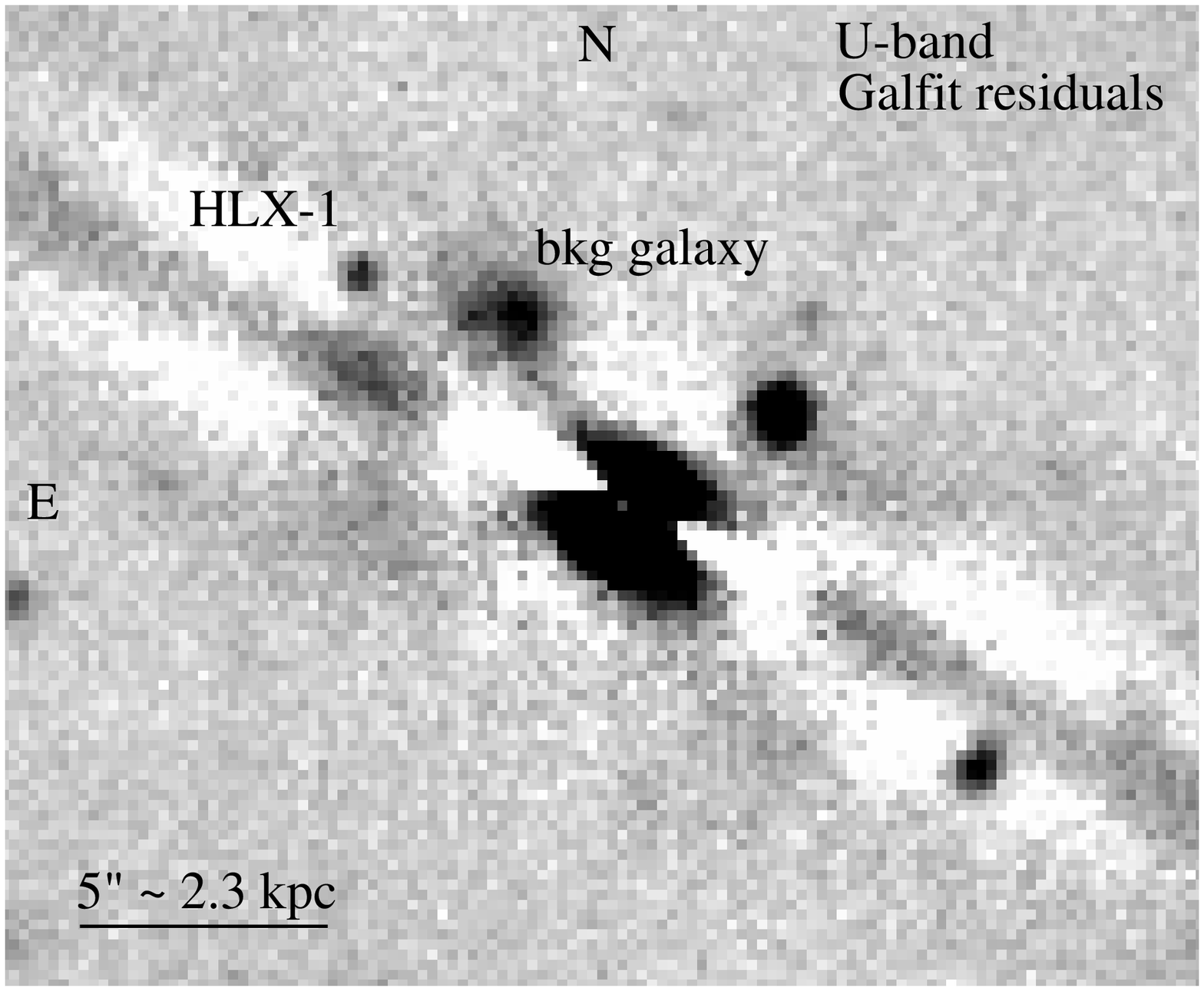,width=84mm,angle=0}
\end{center}
\caption{Top panel: combined U-band image from VLT VIMOS, observed 
on 2010 November 7; in the U-band, HLX-1 is marginally visible by eye 
before background subtraction.
Bottom panel: an example of our efforts to model ESO\,243-49 with several 
{\small{GALFIT}} components (see text).}
\label{f1}
\end{figure}

\begin{figure}
\hspace{-0.5cm}\psfig{figure=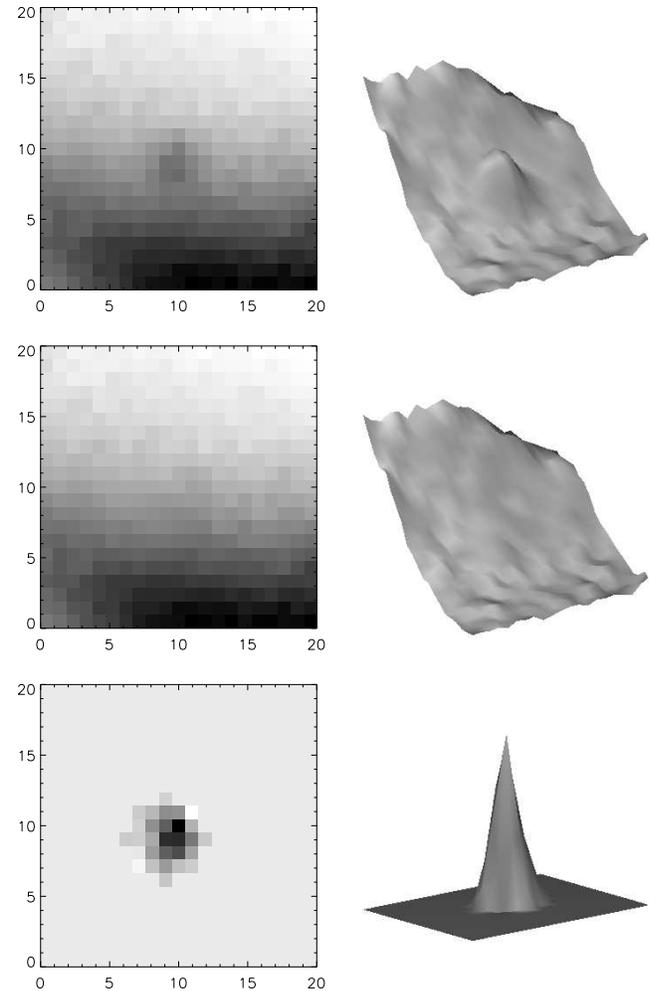,width=97mm,angle=0}
\caption{Top row: the subsection of the VLT VIMOS U-band image around HLX-1. 
Middle row: our best-fitting thin-plate spline local background model. 
Bottom row: the background-model-subtracted U-band image used for 
determining the optical brightness. 
In the left column, axis units are VIMOS pixels (North is up, East is left), 
with 1 pixel $\approx 0\arcsec.2$. In the right column, we display 
the same regions in 3-D format (North at the bottom, East to the right). 
}
\label{f2}
\end{figure}

\begin{table*}
\begin{center}
\begin{tabular}{lcccr}\hline
\hline
Filter  &  Zeropoint & Colour correction & Extinction coefficient & Mean airmass\\
\hline\\[-5pt]
U & $26.09\pm0.11$ & $(0.077\pm0.014)\times(U-B)$  & $0.38 \pm 0.04$ & $1.11$\\
B & $28.19\pm0.02$ & $(-0.056\pm0.005)\times(B-V)$  & $0.24 \pm 0.01$ & $1.21$\\
V & $27.87\pm0.02$ & $(-0.014\pm0.004)\times(B-V)$  & $0.12 \pm 0.01$ & $1.10$\\
R & $28.04\pm0.05$ & $(-0.102\pm0.008)\times(V-R)$  & $0.08 \pm 0.01$ & $1.20$\\
I & $27.64\pm0.04$ & $(0.049\pm0.006)\times(V-I)$  & $0.04 \pm 0.02$ & $1.19$\\
\hline
\end{tabular} 
\end{center}
\caption{VIMOS Q1 zeropoints, colour coefficients, extinction coefficients and 
mean air masses used to convert from count rates to magnitudes.}
\label{tab2}
\end{table*}

\section{Optical data analysis and results}

HLX-1 is clearly detected in emission as a point-like source (Figure 1), 
coincident with the ${\it Chandra}$ position and with the location 
of the optical counterpart identified from our {\it Magellan} images 
last year \citep{sor10}.
In order to estimate its optical flux more accurately, first we had 
to model and remove the contribution of the host galaxy, that dominates 
the background emission particularly at redder colours. To do this, 
we tried two alternative techniques. 

First, we tried modelling the galaxy light profile with {\small {GALFIT}}
\citep{pen02,pen10}, and use the residuals to estimate the emission 
from HLX-1. The galaxy profile is quite complex; we tried fitting it 
with a combination of a S\'ersic profile plus exponential 
disk plus edge-on disk plus disky/boxy profile, and a dust mask.
However, the narrow dust lane across the disk plane creates artifacts 
and ripples that badly affect the wings of the HLX-1 point spread function 
(PSF) and the determination of the background level 
in the surrounding region; much higher spatial resolution 
from {\it HST} imaging \citep{far11} is probably needed for this kind of modelling. 
Nonetheless, at least for the U and B bands, for which the galaxy halo 
contribution is less important, the {\small {GALFIT}} residuals do show 
HLX-1 as a point-like emission source and the background subtraction 
is manageable. About $3\arcsec$ west of HLX-1, 
there is an extended, very blue source, probably a background star-forming 
galaxy \citep{far11}. 
This extended source is mostly responsible for the enhanced, 
asymmetric near-UV emission measured by \citet{sor10} in the north-east 
quadrant of ESO\,243-49. After extensive tests, we concluded that 
the {\small {GALFIT}} method applied to our VLT data provides accurate results 
for the U and B bands, but the background-subtraction error is too large 
for the other bands. We used a 3-pixel source extraction region, 
a background annulus between 5 and 9 pixels, and applied suitable aperture 
corrections for the two bands (based on the PSF of isolated stars) 
to obtain the net count rates within an infinite aperture. We converted them 
to apparent magnitudes as outlined above. We obtained  
$U = 23.84 \pm 0.20$ mag and $B = 25.35 \pm 0.30$ mag.

\begin{figure}
\begin{center}
\hspace{0.5cm}\psfig{figure=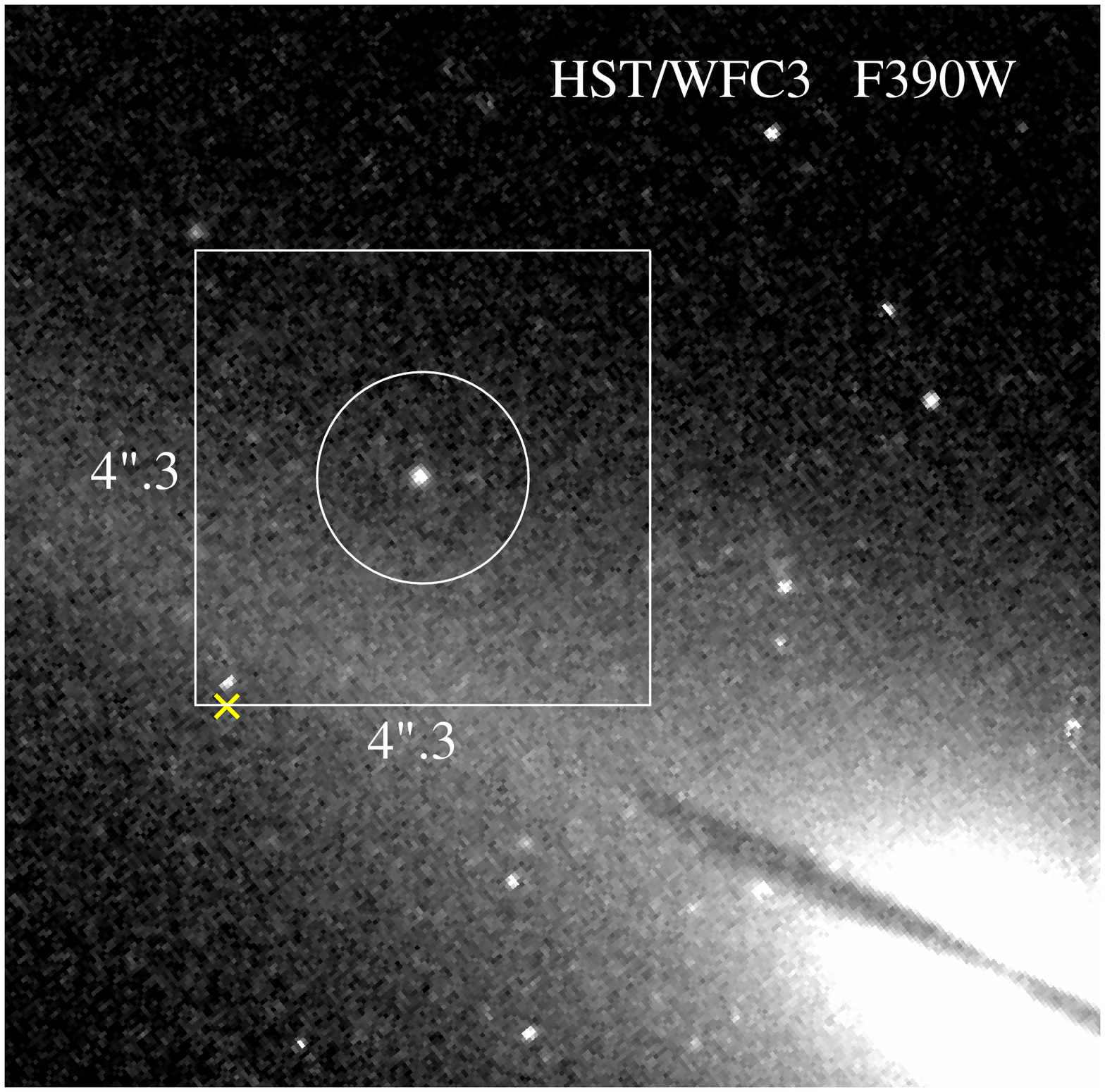,width=67.5mm,angle=0}\\
\hspace{0.5cm}\psfig{figure=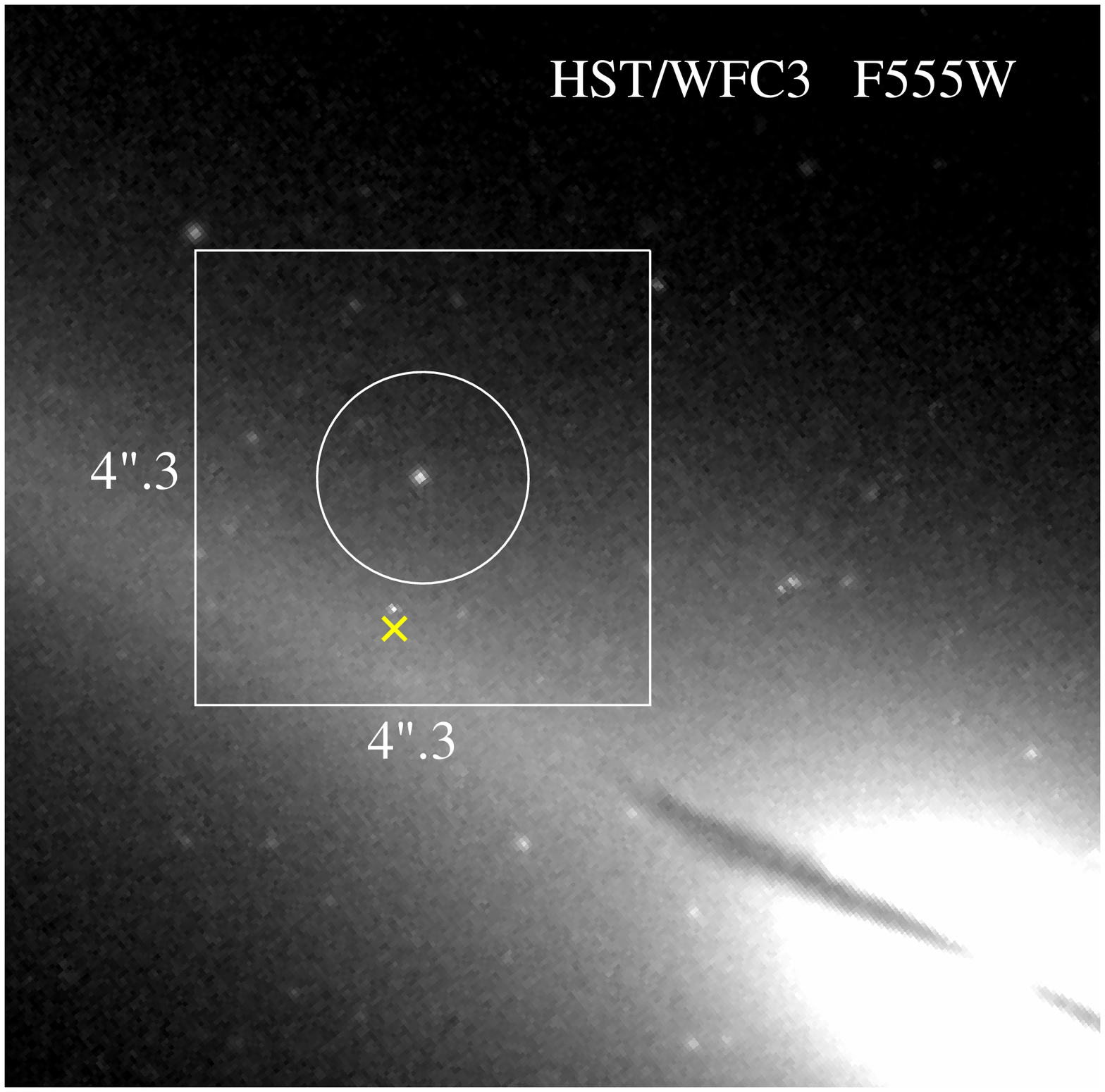,width=67.5mm,angle=0}\\
\hspace{0.5cm}\psfig{figure=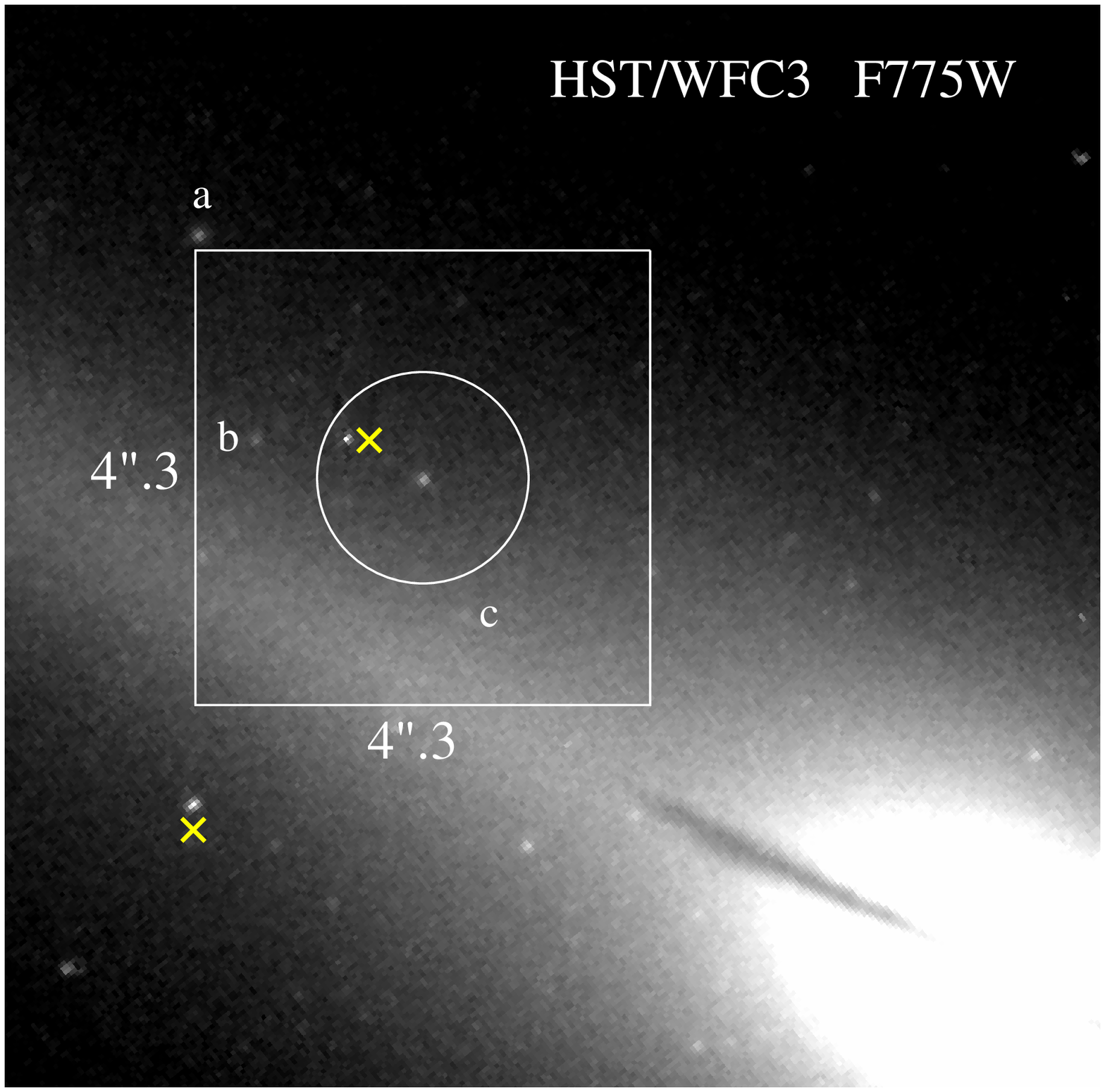,width=67.5mm,angle=0}
\end{center}
\caption{Top panel: {\it HST}/WFC3 image with the F390W filter, 
taken on 2010 September 23, with the outline of the region chosen 
for background modelling in our VLT study. North is up, East to the left. 
Middle panel: same as in the top panel, for the F555W filter image. 
Bottom panel: same, for the F775W filter image. 
In all panels, the box size is $\approx 4\arcsec.3 \times 4\arcsec.3$ 
and the radius of the exclusion circle is $\approx 1\arcsec$. 
Cosmic rays inside or near the extraction regions are marked with yellow 
crosses. 
The three faint sources marked with ``a'', ``b'', ``c'' in the F775W image 
may affect background subtraction in the VLT I band. 
}
\label{f3}
\end{figure}

We then used the following, more accurate technique. We excluded a circle 
of 5-pixel radius around the target and fitted a thin plate spline 
solution to the surrounding $21 \times 21$ (pixel)$^2$ area.
This was done with the {\small{IDL}} routine {\it min\_curve\_surf}. 
This solution effectively follows the background exactly 
outside the exclusion zone, and provides a smooth solution for the background 
inside the exclusion zone (Figure 2).
The main caveat of this technique is that the estimated background within 
the exclusion zone is affected by the values right outside the edge of 
the exclusion zone; in other words, if the exclusion circle is too small, 
the background inside it is overestimated, owing to the wings of HLX-1's 
PSF extending right outside the exclusion circle.
On the other hand, if the exclusion circle is too large, the fitting routine 
does not always provide smooth enough solutions through the whole
aperture. After experimenting with several source and background region sizes, 
we decided to use an exclusion zone radius of 5 pixels and aperture radius 
of 3 pixels, from which we determined the net count rates (subsequently 
corrected to infinite aperture, with aperture corrections 
individually determined for each band). We estimated that the fraction 
of source counts in the PSF wings, falling in the background region, 
outside the 5-pixels exclusion radius, is between 5 and 10 per cent 
across the various filters; this is negligible compared to the number 
of background counts, and does not significantly affect the spline 
fitting or contribute to the error budget. 
We used zeropoint, colour-correction coefficients and extinction coefficients 
as before, to convert our results from count rates to standard (Vegamag) 
magnitudes, and then to AB magnitudes and flux densities (Table 3).
We estimated the errors using a Monte Carlo method: we created synthetic data 
(1000 sets) based on the background model and the measured source count rates. 
Modelling the simulated data in the same way as the real data, we obtained 
the photometric errors quoted in Table 3. 
In summary, our final result is that 
$U = 23.89 \pm 0.18$ mag, $B = 25.19 \pm 0.30$ mag, 
$V = 24.79 \pm 0.34$ mag, $R = 24.71 \pm 0.40$ mag.

HLX-1 was not detected in the I band. This was not surprising, 
given the shorter exposure time and larger unresolved galactic contribution.
We used a Monte Carlo technique 
to estimate the upper limits for the I band flux, in
the following manner: we took the fitted I band background and added 
a fake point-like source of increasing brightness, in order to see 
how bright a source could be ``hidden'' for a given fitted background level 
and measured count rate in the source extraction region.
We found that the $2\sigma$ and $3\sigma$ upper limits 
in the I band are $I>25.8$ mag and $I>24.4$ mag, respectively (Vegamag).

After we had completed our photometry study of the VLT data, 
the set of {\it HST} images obtained and studied by \citep{far11} 
became available in the public archive. We used them to re-examine 
our source extraction 
techniques, checking that there were no other point-like sources in the regions 
we chose for background model fitting, and testing whether 
HLX-1 would still be detectable when blurred to VLT resolution.
We show (Figure 3) that our choice of background regions are indeed 
suitable. While HLX-1 remains easily detectable in the F390W and F555W 
even after smoothing to lower resolutions, regardless of the choice 
of region for background fitting, re-detection of the 
source in the PSF-matched F775W image (the closest match available 
to the I band) depends strongly on background location and 
modelling, which introduces other sources of systematic error.
Moreover, the source was $\approx 1$ mag brighter in every band 
in the {\it HST} observations (Section 5.1). If that applies also 
to the I band, we would expect that $I \approx 25.5$ mag 
in the VLT data; but we see from the {\it HST} F775W image that there are 
another two and possibly three faint sources (marked as ``a'', ``b'', ``c'' 
in the bottom panel of Figure 3) of comparable brightness in I 
($\sim 25.5$--$26$ mag), which may contaminate the background region. 
Such sources would make the background region brighter, and therefore 
we would oversubtract from the HLX-1 source counts\footnote{The same 
contaminating sources are also present in the other bands, but are 
a less significant problem because HLX-1 is brighter.}.
For these reasons, we list conservative non-detection 
upper limits to the I band.
We stress that the other four bands 
with robust detections already provide the information needed for 
the main conclusions of our study.


\begin{table*}
\begin{center}
\begin{tabular}{lccccccc}\hline
\hline
Filter  & $\lambda_{\rm eff}$ &  $m$ & $m_{\rm AB}$  & $M_0$ 
 & $F^{\rm obs}_\lambda$  & $F^0_\lambda$ & $F^1_\lambda$ \\[5pt]
 & (\AA) & (Vegamag) & (ABmag)  & (Vegamag) 
 &  (erg cm$^{-2}$ s$^{-1}$ \AA$^{-1}$) &
 (erg cm$^{-2}$ s$^{-1}$ \AA$^{-1}$) & (erg cm$^{-2}$ s$^{-1}$ \AA$^{-1}$) 
\\
\hline\\[-5pt]
U & $3650$ & $23.89\pm0.18$  & $24.67\pm0.18$ & $-11.07\pm0.18$
       & $(11.1^{+2.1}_{-1.7}) \times 10^{-19}$ 
       &  $(11.8^{+2.2}_{-1.8}) \times 10^{-19}$ 
       &  $(14.5^{+2.6}_{-2.1}) \times 10^{-19}$  \\[5pt]
B & $4380$ & $25.19\pm0.30$   & $24.99\pm0.30$ & $-9.76\pm0.30$
       & $(5.7^{+1.9}_{-1.4}) \times 10^{-19}$ 
       &  $(6.1^{+2.0}_{-1.5}) \times 10^{-19}$ 
       &  $(7.1^{+2.4}_{-1.7}) \times 10^{-19}$   \\[5pt]
V & $5450$ & $24.79\pm0.34$  &  $24.81\pm0.34$ & $-10.14\pm0.34$
       & $(4.4^{+1.6}_{-1.3}) \times 10^{-19}$ 
       &  $(4.5^{+1.7}_{-1.3}) \times 10^{-19}$ 
       &  $(5.2^{+1.9}_{-1.5}) \times 10^{-19}$   \\[5pt]
R & $6410$ & $24.71\pm0.40$ &  $24.87\pm0.40$ & $-10.22\pm0.40$
       & $(3.0^{+1.4}_{-1.0}) \times 10^{-19}$ 
       &  $(3.1^{+1.4}_{-1.0}) \times 10^{-19}$ 
       &  $(3.4^{+1.6}_{-1.1}) \times 10^{-19}$   \\[5pt]
I & $7980$ & $>24.4$  & $>25.0$ & $>-10.0$ & $<1.7 \times 10^{-19}$ 
       &  $<1.8 \times 10^{-19}$ &   $<1.9 \times 10^{-19}$ \\[5pt]
\hline
\end{tabular} 
\end{center}
\caption{Summary of observed and de-reddened magnitudes and flux densities 
for HLX-1 during our 2010 November observations. $m$ and $m_{\rm AB}$ are the  
apparent magnitudes at the top of the atmosphere (Vegamag and ABmag 
units, respectively). $M_0$ is the intrinsic absolute magnitude, 
corrected for a line-of-sight Galactic extinction $A_V = 0.04$ mag 
($E(B-V) = 0.013$ mag) 
and a distance modulus $=34.89$ mag. $F^{\rm obs}_\lambda$ is the observed 
top-of-the-atmosphere flux density ({\it i.e.}, the flux density 
corresponding to $m_{\rm AB}$). $F^0_\lambda$ is the flux density corrected 
for a line-of-sight Galactic extinction $A_V = 0.04$ mag (a likely 
lower limit to the total optical extinction). 
$F^1_\lambda$ is the flux density corrected 
for an extinction $A_V = 0.18$ mag (a likely 
upper limit to the total optical extinction).
The upper limits to the (non-detected) $I$ brightness and flux are at the 99\% 
confidence level. The other error ranges are $1\sigma$.} 
\label{tab3}
\end{table*}



We converted from observed brightnesses in the Vegamag system 
to top-of-the-atmosphere flux densities 
using VIMOS transmission models available online\footnote{http://www.eso.org/observing/etc}. The VIMOS U-band filter is moderately different \citep{non09}
from the Johnson U-band filter \citep{bes90} used to define the standard 
photometric system, while the other VIMOS filters have standard passbands.
We double-checked our results using standard Vegamag-to-ABmag 
conversion 
tools\footnote{http://www.stsci.edu/hst/nicmos/tools/conversion\_form.html}.
In all cases, we adopted a blackbody input spectrum at $T=25,000$ K 
for the flux density conversion (Table 3), but we also calculated 
the flux densities in the case of a flat spectrum, a $T=10,000$ K 
blackbody, and various stellar models from O to A type. 
The reason we adopted $T=25,000$ K is that the source has blue 
U-B and U-V colours, and does not appear to have a Balmer absorption jump 
typical of cooler stellar atmospheres. Anyhow, in all bands 
except for B, the flux density at the effective wavelength of the filter 
(also known as pivot wavelength), for a given magnitude, is essentially 
independent of spectral shape, within $\sim 1\%$. There is a small dependence 
on spectral shape/stellar type for the B band, because the Balmer jump falls 
at the high-frequency end of that passband. This introduces a systematic 
error of $\approx \pm 15\%$ of the flux values listed for the B band 
in Table 3, which is in any case less than the photometric uncertainty 
in the measured brightness.

Finally, we corrected for the optical extinction, in order to determine 
the true emitted luminosity and colours.
As a first approximation, we took the foreground extinction 
$E(B-V) = 0.013$ mag (corresponding to $A_V \approx 0.04$ mag) 
from \citep{sch98}.
An alternative possibility is to take the maps of Galactic HI 
from \citet{kal05}, and convert $N_{\rm H}$ to an optical extinction
for example through the relation of \citet{guv09}, 
$N_{\rm H} = (2.21 \pm 0.09) \times 10^{21} A_V$ 
(corresponding to $A_V \approx 0.08$ mag).
There may be additional extinction components, intrinsic to ESO\,243-49, 
and in the local environment of HLX-1. For the ESO\,243-49 component, 
we note that the halo of an S0 galaxy dominated by stellar populations 
older than 1 Gyr is not expected to contain much gas and dust; 
indeed, it was found \citep{dis03} that the best-fitting column densities  
for the brightest X-ray sources in the Sombrero galaxy 
(an edge-on S0 galaxy similar to ESO\,243-49) are consistent or only marginally 
higher than the Galactic line-of-sight value.
Moreover, optical/X-ray studies of Galactic black hole transients 
(fed via Roche-lobe overflow, as is likely the case in HLX-1) 
have shown \citep{hyn05} that the measured optical extinction 
is always smaller or comparable to the value that would be inferred 
from the values of $N_{\rm H}$ fitted to the X-ray spectra.
This is because the local gas that contributes to the X-ray absorption 
is mostly confined to the X-ray emitting region, and/or is too hot to include 
a dusty component, so it does not contribute to the optical/UV absorption.
In the case of HLX-1, we found from the {\it Swift}/XRT spectra 
that the total $N_{\rm H}$ fitted to the X-ray source is consistent 
with the foreground value, or is at most twice that value (Table 4).
This is the same result obtained by \citet{ser11} (total 
$N_{\rm H} = (3\pm1) \times 10^{20}$ cm$^{-2}$ in the thermal state), 
while \citet{god09} adopt $N_{\rm H} \approx 4 \times 10^{20}$ cm$^{-2}$.
Since we argued that the X-ray absorption provides an upper limit 
to the optical extinction, $A_V \la 0.18$ mag. In Table 3, we list 
the top-of-the-atmosphere fluxes, those corrected with $A_V = 0.04$ mag, 
and those for $A_V = 0.18$ mag. The difference is small enough 
that none of our conclusions depends on the precise value 
of the extinction.


\section{X-ray irradiation}

One of our main objectives is to determine what fraction of 
the optical luminosity is due to re-emission of intercepted X-ray photons.
Thus, we have to monitor the X-ray luminosity during the epoch of our VLT 
observations, and compare it with X-ray and optical luminosities at other 
epochs. Fortunately, HLX-1 has been the target of more than 100 {\it Swift} 
X-ray Telescope (XRT) observations since 2008 October; see NASA's HEASARC 
data archive for a detailed logbook. We used the on-line XRT data product 
generator \citep{eva07,eva09} to extract light curves and spectra 
(including background and ancillary response files); we selected grade 
$0$--$12$ events. We downloaded the suitable spectral response file 
for single and double events in photon-counting mode from the latest 
{\it Swift} Calibration Database. 

In 2010 November, HLX-1 was in the decline phase after the 2010 X-ray 
outburst. The X-ray luminosity did not change significantly between 
2010 November 7 and 2010 November 26 (Figure 4), and the timescale 
for the exponential decay is much longer than 19 days. Thus, 
we can assume that the irradiation correction to the $R$ band 
(observed on November 26) should not be significantly different 
from that affecting the optical bands on November 7.
The individual {\it Swift}/XRT observations around the two epochs are 
too short to allow for meaningful spectral analysis. Therefore, to characterize 
the X-ray spectrum, we coadded all the observations in the middle part 
of the outburst decline, between 2010 September 30 and 2010 December 11, 
a period in which the X-ray source was almost on a plateau.  
Moreover, we coadded all {\it Swift}/XRT observations from 2008 October
to 2011 August 
in which the $0.3$--$10$ keV count rate was between 0.01 and 0.02 
ct s$^{-1}$. We then fitted both co-added spectra with {\small {XSPEC}} 
version 12 \citep{arn96}, using the Comptonized, irradiated disk model 
{\it {diskir}} \citep{gie08}.

An advantage of the {\it {diskir}} model over widely used two-component 
models such as {\it {diskbb}} plus {\it {power-law}} is that the Comptonized 
component is physically truncated at low energies, and the disk spectrum 
can easily be extrapolated to the optical bands, and compared with 
the flux from the optical counterpart.
We fixed the fraction $f_{\rm in}$ of luminosity in the Compton tail that is 
thermalized in the inner disk to $0.1$ \citep{gie08}. The other 
irradiation parameter in the model, $f_{\rm out}$, is the fraction 
of bolometric flux that is thermalized in the outer disk. The value of 
$f_{\rm out}$ and of the outer disk radius $r_{\rm out}$ are essentially 
unconstrained when we fit the X-ray data, especially for a cool disk 
with peak temperature $\la 0.2$ keV, but will become important 
for the optical extrapolation.

We find that the long- and short-baseline {\it Swift}/XRT spectra are 
essentially identical in shape and normalization (Figure 5). 
The exposure time for the long spectrum was $1.2 \times 10^5$ s; 
the exposure time for the short spectrum (which is simply a subset of 
the longer-baseline data) was $4.7 \times 10^4$ s.
The best-fitting parameters for the long-baseline spectrum 
are listed in Table 4. This result is consistent with the expected spectral 
behaviour of HLX-1 at intermediate luminosities \citep{ser11}, 
with a soft thermal component that is still dominant, and an emerging 
power-law that becomes harder during the decline.
The emitted luminosity $L_{0.3-10 {\rm keV}} \approx 5.5 \times 10^{41}$ 
erg s$^{-1}$, a factor of 2 lower than at the outburst peak.
The disk normalization implies an inner radius 
$r_{\rm in} \approx 6 \times 10^4$ km (for an approximately face-on disk), 
indicative of a BH mass $\sim 10^4 M_{\odot}$ and consistent with previous 
X-ray studies \citep{dav11,ser11}.

\begin{figure}
\begin{center}
\psfig{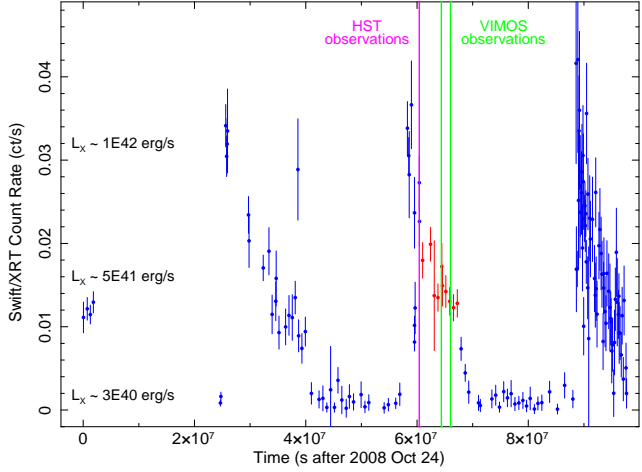}
\end{center}
\caption{{\it {Swift}}/XRT lightcurve in the $0.3$--$10$ keV band, 
between 2008 October 24 and 2011 November 28. The two epochs of our 
VLT observations are indicated by vertical green lines; the epoch 
of the {\it HST} observations is also plotted.
We integrated to {\it {Swift}}/XRT spectrum from the epochs plotted in red, 
to determine the irradiation conditions around the time of our 
VLT observations.}
\label{f4}
\end{figure}

\begin{figure}
\begin{center}
\psfig{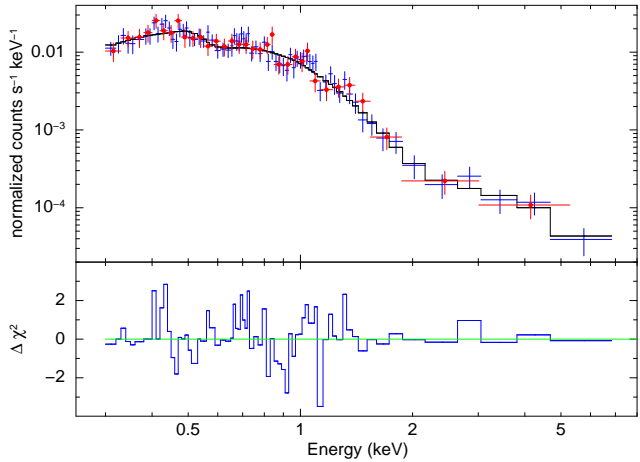}
\end{center}
\caption{{\it {Swift}}/XRT spectrum modelled with the Comptonized, 
irradiated disk model {\it {diskir}} in {\small{XSPEC}}. The blue spectrum 
includes all the epochs when the XRT count rate was between 0.01 and 0.02 
ct s$^{-1}$. The red datapoints are a subset of the blue datapoints, 
for the epochs plotted in red in Figure 4. The $\chi^2$ residuals 
in the bottom panel refer to the long-baseline spectrum, 
which has better signal-to-noise and provides stronger constraints.}
\label{f5}
\end{figure}

\begin{table}
\begin{center}
\begin{tabular}{lr}
\hline
Parameter &  Value  \\
\hline\\[-5pt]
$N_{\rm H,Gal}$ & $[2.0]$  \\[5pt]
$N_{\rm H,int}$ & $0.8^{+2.4}_{-0.8}$\\[5pt]
$kT_{\rm in}$ & $0.17^{+0.01}_{-0.05}$\\[5pt]
$\Gamma$ & $1.59^{+0.68}_{-0.34}$\\[5pt]
$kT_{e}$ & $[100]$\\[5pt]
$L_{\rm C}/L_{\rm d}$ & $0.60^{+4.33}_{-0.45}$\\[5pt]
$f_{\rm in}$ & $[0.1]$\\[5pt]
$r_{\rm irr}$ & $[1.2]$\\[5pt]
$f_{\rm out}$ & $[2\times10^{-3}]$\\[5pt]
$\log (r_{\rm out}/r_{\rm in})$ & $[3.45]$\\[5pt]
$N_{\rm disk}$ & $40.6^{+91.7}_{-13.5}$\\[5pt]
\hline\\[-5pt]
$f_{0.3-10}$ & $4.1^{+0.2}_{-0.2}$ \\[5pt]
$L_{0.3-10}$ & $5.5^{+0.4}_{-0.6}$\\[5pt]
\hline\\[-5pt]
$\chi^2_{\nu}$ & $0.83~(55.3/67)$\\[5pt]
\hline
\end{tabular} 
\end{center}
\caption{Best-fitting spectral parameters for the coadded 
{\it Swift}/XRT observations during intermediate-luminosity states. 
Units: $N_{\rm H,Gal}$ and $N_{\rm H,int}$ are in units of $10^{20}$ cm$^{-2}$; 
$kT_{\rm in}$ and $kT_{e}$ in keV; $r_{\rm irr}$ in units of the inner-disk 
radius $r_{\rm in}$;
$f_{0.3-10}$ in $10^{-13}$ erg cm$^{-2}$ s$^{-1}$; 
$L_{0.3-10}$ in $10^{41}$ erg s$^{-1}$. The disk normalization 
$N_{\rm disk} = \left[\left(r_{\rm in}/{\rm km}\right)/\left(D/{\rm 10kpc}\right)\right]^2 \cos \theta$, where $\cos \theta$ is the viewing angle of the disk.
Errors indicate the 90\% confidence interval for each parameter of interest.}
\label{tab4}
\end{table}

\section{Optical/X-ray comparison}

After measuring the optical brightness in the VLT dataset, 
and modelling the temperature 
and luminosity of the X-ray-emitting accretion disk, we shall 
now discuss whether the optical emission is more likely to come 
from the outer part of the irradiated disk, from a stellar population 
or from a single donor star. A comparison with the optical brightness 
in the {\it HST} observations \citep{far11} provides a crucial test 
to distinguish between those models.

\subsection{Optical decline between the {\it HST} and VLT observations}

The X-ray luminosity, as determined from the {\it Swift}/XRT count rate 
and spectra, was a factor of $\approx 2$ higher during the 2010 September 23 
{\it HST} observations than during our VLT observations, taken at a later 
stage of the 2010 outburst decline. 
We showed (Table 3) that the optical/near-UV brightness of HLX-1 
was $\approx 24.7$--$25.0$ mag (ABmag system) in out VLT data; 
in the {\it HST} observations, it was $\approx 23.8$--$24.0$ mag 
\citep{far11}\footnote{We also independently checked that HLX-1 
was significantly brighter in the {\it HST} images, by doing relative 
photometry with a dozen other sources detected in both fields.}.
Although the VIMOS filter bands are slightly different from the 
{\it HST} bands, this discrepancy in flux density occurs in every band 
and cannot be explained with colour transformations; as we are referring 
to the top-of-the-atmosphere flux density, this is also independent 
of the different choices of line-of-sight and intrinsic extinction 
corrections. Interpolating 
the flux densities given by \citet{far11} across our observed bands, 
we determine that the VLT flux was fainter than during the {\it HST} 
observations, with $3\sigma$ significance in each of the U,B,V and I bands, 
and $2.5\sigma$ significance in R.
Thus, we conclude that the source had significantly dimmed, 
by $\approx 1$ mag, in the near-UV/optical/near-IR bands. 
To test possible further evidence of variability in the optical counterpart, 
we also re-examined our Magellan observations \citep{sor10}, taken 
a few days after the 2009 August outburst peak, when the X-ray flux was, 
again, a factor of 2 higher than in 2010 November. 
The Magellan R-band brightness $m_{{\rm AB}, R} \approx 24.0 \pm 0.3$ mag  
is $\approx 0.9$ mag higher than in our 2010 VLT observations, 
with a $2\sigma$ significance. 
(The larger error in the Magellan V-band measurement, which we now 
re-estimate as $m_{{\rm AB}, V} \approx 24.5 \pm 0.5$ mag prevents 
firm conclusions on that band.)

On its own, our discovery of long-term optical variability correlated with 
the X-ray luminosity already rules out the possibility that 
the optical/UV counterpart was dominated by emission from 
an old globular cluster or a young super-star-cluster 
during the {\it HST} observations, as discussed in \citet{far11}.
Hence, we suggest that the dominant optical/UV contribution 
near the outburst peak was from an irradiated accretion disk.
Any additional contribution from a (constant-luminosity) stellar population 
must be fainter than the luminosity determined from our VLT observations.

In fact, even in our VLT data, the accretion disk is likely to give 
a significant contribution to the optical/UV flux, as the X-ray 
luminosity was still 20 times higher than in the low state, 
and the soft X-ray flux was still dominated by (partly Comptonized) 
disk emission (Table 4).
In the rest of this Section, we are going to discuss and constrain 
the possible contribution of disk and stellar populations 
in our VLT data.

\subsection{Irradiated disk}

The simplest scenario is a standard \citep{sha73} non-irradiated disk, 
with inner temperature and normalization derived from X-ray fitting (Table 4), 
extrapolated to an infinite radius (for practical purposes, 
$r_{\rm out} = 10^5 r_{\rm in}$).
This is plotted as a solid black line in all panels of Figure 6; 
in those plots, X-ray absorption has been removed from the disk model, 
and the optical datapoints have been corrected for a line-of-sight 
reddening $E(B-V) = 0.013$ mag\footnote{Using our upper limit 
$E(B-V) = 0.06$ mag does not change any of the arguments in this Section.}.
In the optical band, the predicted non-irradiated disk emission 
approximately reproduces the spectral slope between U and R but falls 
a factor of 3 below the observed brightnesses.

Clearly, irradiation of the outer disk is needed to increase the optical 
flux of the model without changing the X-ray luminosity. 
First, we assumed an irradiation fraction $f_{\rm out} = 10^{-3}$ 
while keeping a large outer radius ($r_{\rm out} = 10^5 r_{\rm in}$). 
The predicted optical flux (solid green line in Figure 6) 
is now roughly consistent with the VLT data from U to R (although 
the U-V colour is too red), but is much too bright for the I band. 
As a general rule, the optical emission of a large irradiated disk 
is redder than the emission of a non-irradiated disk. However, 
our source is significantly bluer.

Truncating (or shading) the outer edge of the disk removes 
some of the red emission from the outer regions, and makes the optical 
slope steeper. 
We find that the optical data are well fitted (solid blue line in Figure 6)
with emission from the same accretion 
disk seen in the X-ray band, extending to $r_{\rm out} \approx
2800 r_{\rm in} \approx (1.7 \times 10^{13})/\cos \theta$ cm,
and with an intercepted and re-radiated fraction of X-ray photons 
$f_{\rm out} \approx 2 \times 10^{-3}$, a very plausible value 
for irradiated disks in binary systems. 
Reprocessing fractions of $\sim$ a few times $10^{-3}$  
are suggested both by theoretical modelling 
\citep[e.g.,][]{vrt90,dej96,kin98,dub99} and by observations 
of accretion disks in Galactic BHs 
\citep[e.g.,][]{hyn02,gie09}.
In summary, 
considering that an irradiated disk was the dominant optical/UV emitter 
in 2010 September (Section 5.1), and that the X-ray flux from the inner disk 
was still relatively high in 2010 November, we take the irradiated disk 
scenario as the most likely interpretation also for the optical 
emission in our VLT data.

\begin{figure}
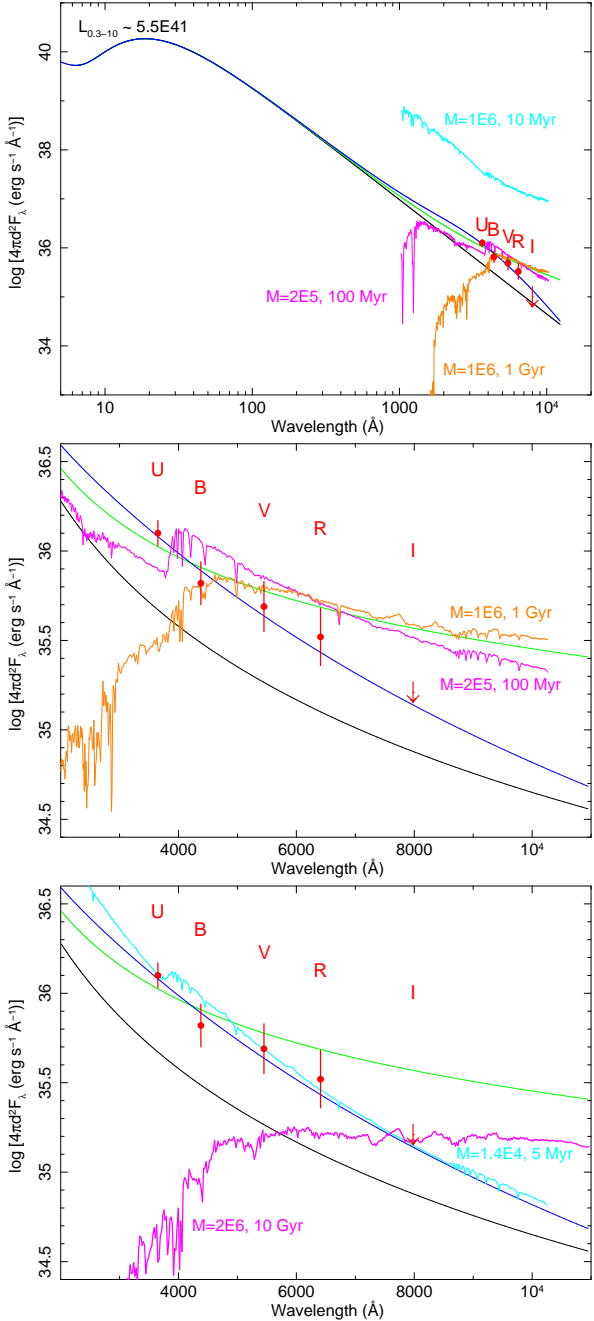

\begin{center}
\hspace{0.3cm}\psfig{figure=fig6a.ps,width=78mm,angle=270}\\
\hspace{0.3cm}\psfig{figure=fig6b.ps,width=78mm,angle=270}\\
\hspace{0.3cm}\psfig{figure=fig6c.ps,width=78mm,angle=270}
\end{center}
\caption{Top panel: optical/UV flux densities 
from our VLT data (plotted in red), compared with 
the model flux density predicted from various types of accretion disks 
and stellar populations. At high energies, the disk parameters 
are fixed from the fit to the {\it {Swift}}/XRT data. 
Black line: standard, non-irradiated disk spectrum; green line: irradiated 
disk; blue line: irradiated, truncated disk; cyan line: 
spectrum of a young, massive star cluster; magenta line: 
spectrum of an intermediate-age stellar population; 
orange line: spectrum of an old globular cluster.
Middle panel: zoomed-in view in the optical/UV band, plotted in linear scale: 
the three disk spectra and the two old/intermediate-age stellar populations 
are as in the top panel.
Bottom panel: the spectrum 
of a low-mass, very young stellar population (cyan line) fits the data 
as well as the irradiated, truncated disk model (blue line) but may not 
have a simple physical interpretation; an very old, massive globular cluster 
may also be associated with HLX-1, provided that its mass is $\la 2 \times 
10^6 M_{\odot}$ (magenta line). }
\label{f6}
\end{figure}

\begin{figure}
\begin{center}
\psfig{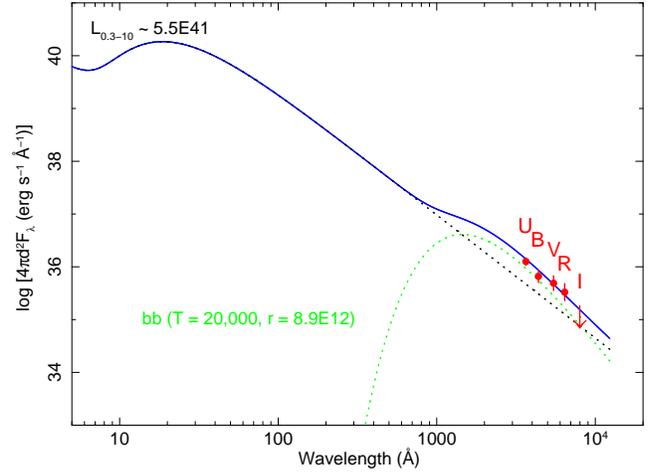}
\end{center}
\caption{Optical/UV flux densities from our VLT data (plotted in red), 
compared with a model consisting of a standard non-irradiated disk (dotted 
blue line) plus a single blackbody with $T = 20,000$ K 
and characteristic size $\approx 8.9 \times 10^{12}$ cm $ \approx 130 R_{\odot}$ 
(dotted green line).}
\label{f7}
\end{figure}

\subsection{Stellar population}

An alternative possibility is that most of the optical 
emission in the VLT data comes from a massive star cluster surrounding the BH.
For example, this is what we expect if the IMBH was the nucleus 
of a recently accreted satellite dwarf, or was formed from core collapse 
in a massive star cluster.
We ran instantaneous star formation simulations with {\small {STARBURST99}} 
\citep{lei99,vaz05} to determine the expected 
brightness and colours of star clusters of various masses and ages 
(we assumed solar metallicity for simplicity, and a Kroupa initial mass 
function). Here, we focus on a few representative cases.

We find that a young super-star-cluster (for example, the solid cyan line 
in the top panel of Figure 6 represents a cluster with $M = 10^6 M_{\odot}$ 
and age of 10 Myr) is inconsistent with the VLT optical data: 
it would have blue colours and small or no Balmer jump, consistent 
with the observed data, but would be much too bright. This rules out 
one of the scenarios presented in \citet{far11}. 
A young star cluster with much lower mass 
can match the observed optical luminosity and colours; 
the solid cyan line in the bottom panel of Figure 6 
represents a cluster with $M = 1.4 \times 10^4 M_{\odot}$ 
and age of 5 Myr. In fact, this is only 
an upper limit to the young cluster mass, if there is still a significant 
disk contribution (Section 5.2).
A stellar population age $\la 6$ Myr is required in order to match 
the optical colours, at solar metallicity---modulo the age-metallicity 
degeneracy\footnote{In simple terms, the age-metallicity 
degeneracy means that colours and luminosity 
are the same for a stellar population with half the metal abundance 
and three times the age \citep{wor94}.}.
However, the physical interpretation of this scenario 
is problematic. A $10^4 M_{\odot}$ cluster is too small to harbour
an IMBH formed from runaway core collapse \citep{por02,fre06}, 
especially one as massive as required for HLX-1 \citep{dav11}, 
similar to the mass of the cluster itself.
Its very young age makes it unlikely that 
the cluster could be much more massive at birth.  
Moreover, the {\it HST} near-UV images show no other signs of ongoing 
or very recent, clustered star formation in the region surrounding HLX-1, 
which appears like a one-off object.
For all these reasons, we conclude that the young-star-cluster scenario 
is much less likely than the irradiated disk scenario.

Old globular clusters are much fainter for a given stellar mass. 
A cluster with $M = 10^6 M_{\odot}$ and age of 1 Gyr would be marginally 
consistent with the observed B, V, R brightnesses (solid orange line 
in Figure 6, middle panel), 
but totally inconsistent with the observed colour $U-B \approx -1.3$ mag, 
which is typical of OB stars, and with the upper limit to the I band.
Intermediate-age ($\approx 50$--$100$ Myr) clusters with masses 
$\sim 10^5 M_{\odot}$ are a slightly better match to the B, V, R optical 
data but remain inconsistent with U and I (solid magenta line in Figure 6, 
middle panel). The strongest constraint against intermediate-age and 
old stellar populations comes from the apparent lack of a Balmer jump 
between B and U, in our VLT observations; 
this implies that Hydrogen is mostly ionized at the photosphere 
of the optical emitter. If the emitter is a stellar population, 
the lack of a Balmer jump alone strongly suggests an age $\la 50$ Myr. 
We conclude that the optical counterpart of HLX-1 cannot be 
dominated by a massive old or intermediate-age globular cluster. 

More complex and/or contrived theoretical scenarios can also be proposed. 
For example, the IMBH could be located in an old, faint globular cluster 
(perhaps the parent cluster that formed the IMBH via core collapse 
in earlier epochs) with an age $> 1$ Gyr and current mass 
$\la$ a few $\times 10^5 M_{\odot}$ (thus, satisfying the VLT 
luminosity constraints), which has recently undergone a new burst 
of star formation, that produced a $\sim 10^4 M_{\odot}$ young stellar 
population (including perhaps the donor star to the IMBH). 
Or, the blue optical emission may come entirely from the IMBH disk, 
inside an old globular cluster, and the mass donor may be an evolved 
low-mass star. For a globular cluster with an age of 10 Gyr, 
the upper mass limit consistent with the VLT data is 
$\approx 2 \times 10^6 M_{\odot}$ (Figure 6, bottom panel). 
Such an old cluster would dominate the near-IR emission but 
would give a negligible contribution to the optical/UV bands, 
compared with the irradiated disk and/or young stellar population.
We do not have enough observational data 
at the moment to test such multiple-component scenarios.
New observations of HLX-1 during a much lower X-ray state, and in the IR, 
will provide conclusive tests between the irradiated disk and star cluster 
models, and may provide a more quantitative estimate or upper limit 
to the optical contribution of a stellar population around the BH.



Finally, we may want to consider the possibility that the optical 
emission comes from the surface of an individual donor star.
The inferred absolute magnitude $M_V \sim M_B \sim -10$ mag 
is very high but not beyond the limit of existing stars, 
for example blue hypergiants and Luminous Blue Variables (LBVs).
Stars in this luminosity range include Cyg OB2-304, the most luminous 
star known to-date in the Galaxy \citep[$M_V = -10.6$ mag:][]{mas01}; 
$\eta$ Carinae was also at $M_V \la -10$ mag 
in the early 19th century, before the 1843 Great Eruption \citep{fre04}. 

As an extreme case, we modeled a standard non-irradiated disk 
(based on the X-ray parameters, as in Section 5.1) with 
an additional single-temperature blackbody component.
We find that for $T_{\rm bb} \approx 20,000$ K and a stellar radius 
$r_{\ast} \approx 8.9 \times 10^{12}$ cm $\approx 130 R_{\odot}$, 
the resulting spectrum also matches the observed optical data (Figure 7) 
as well as the irradiated disk spectrum. We cannot distinguish 
between the two cases with the scant optical data at hand.
What this means is that most of the optical emission may come 
from a region with characteristic size $\sim 10^{13}$ cm 
and characteristic temperature $\sim 20,000$ K: either the outer radius 
of the irradiated (part of the) disk, or the irradiated face 
of a large, massive donor star.
A scenario where the optical luminosity is dominated by a single LBV 
or blue hypergiant donor star implies the presence of ellipsoidal 
variations, due to the tidally-distorted shape of the star,  
and orbital modulations, due to the variable fraction of 
the irradiated face of the star visible to us during the orbit. 
The single-donor scenario may perhaps explain the outburst period 
$\approx 370$ d seen in the X-ray lightcurve \citep{las11}, 
if the orbit is elongated and Roche-lobe mass 
transfer occurs mostly at periastron, with the formation of a disk. 
However, it is very unlikely that the optical/UV counterpart 
is a single donor star, because there are no other similar objects 
or other evidence of exceptional star formation in the surrounding regions, 
within at least 1 kpc. Another issue to consider is that such massive stars 
are shrouded in dense winds, in apparent contrast with the low column 
density measured from the X-ray spectra; on the other hand, the wind 
may be completely ionized because of the X-ray emission from the BH.
Deeper optical spectra of the optical counterpart may help test between 
those scenarios.

\section{Conclusions}

We observed the optical counterpart of HLX-1 with VLT VIMOS, in 2010 November, 
when the X-ray source was in an intermediate state, declining 
after the 2010 August outburst. 
We found the following brightness values in the standard Vegamag system: 
$U = 23.89 \pm 0.18$ mag; $B = 25.19 \pm 0.30$ mag; 
$V = 24.79 \pm 0.34$ mag; $R = 24.71 \pm 0.40$ mag.
In the ABmag system, this corresponds to 
$m_{\rm{AB},U} = 24.67 \pm 0.18$ mag, $m_{\rm{AB},B} = 24.99 \pm 0.30$ mag, 
$m_{\rm{AB},V} = 24.81 \pm 0.34$ mag, $m_{\rm{AB},R} = 24.87 \pm 0.40$ mag.
We do not detect the source in the I band ($m_{\rm{AB},I} > 25.0$ mag); 
this was expected because the I band had the shortest exposure time 
and the largest galactic contamination.

We found that the optical/UV brightnesses in every band 
are $\approx 1$ mag fainter 
than in the {\it HST} observations from 2010 September, 
at a time when the X-ray flux was a factor of 2 higher. 
This rules out the scenario \citep{far11} that the optical source 
seen in the {\it HST} observations was dominated by a massive star cluster, 
because in that case its brightness would not have changed.
It strongly suggests that the optical counterpart was dominated 
by an irradiated accretion disk.

From the VLT data alone, we cannot tell whether the optical emission 
was still dominated by an irradiated disk, or by a stellar population: 
optical observations at even lower X-ray flux levels will 
test these alternatives. We can, however, constrain 
the maximum stellar mass and age of the stellar population that can be 
consistent with the observed VLT luminosities and colours. 
We found that they are consistent with a young stellar population 
of mass $\sim 10^4 M_{\odot}$ and age $\la 10$ Myr. This immediately 
rules out a young super-star-cluster, which would have similar 
blue colours and lack of a Balmer jump, but would be much too luminous.
If there is an old globular cluster around HLX-1, as previously 
speculated \citep{sor10},
it may dominate the near-IR bands, but it cannot give 
a significant contribution to the observed optical/UV flux 
discussed in this work. We estimate that 
a 10-Gyr-old globular cluster must have a mass $\la 2 \times 10^6 M_{\odot}$ 
to be consistent with the VLT data. By comparison with the current 
models and observational mass limits on IMBH candidates in old globular 
clusters \citep[Table 1 in][]{van10}, we argue that a host cluster 
mass $\approx 2 \times 10^6 M_{\odot}$ may be just high enough 
at least to allow for the possibility of a central IMBH 
with a mass $\approx 10^4 M_{\odot}$, as suggested by X-ray modelling 
\citep{ser11,dav11}. 

In conclusion, we attribute most of the optical/UV emission to an irradiated 
disk, as the simplest scenario consistent with all the data at hand. 
The more complex scenario of a very old globular cluster with a small 
burst of recent star formation is also acceptable, but requires 
more observational data to be tested.
We modelled the X-ray spectrum around the time of the VLT observations 
(when $L_{0.3-10} \approx 5.5 \times 10^{41}$ erg s$^{-1}$), 
and during epochs of similar levels of X-ray luminosity in previous years.
We showed that an irradiated disk can explain the observed optical 
luminosity and colours, with a plausible reprocessing fraction 
$\approx 2 \times 10^{-3}$, typical of irradiated disks in X-ray binaries.
Regardless of any details on disk structure and irradiation, 
if the optical emission is thermal, during the VLT observations 
it comes from a region with characteristic size $\sim 10^{13}$ cm 
and characteristic temperature $\approx 20,000$ K; this could be the size 
and temperature of the irradiated outer disk. 







\section*{Acknowledgments}

We thank Chris Done, Sean Farrell, Paul Kuin, Kip Kuntz, Manfred Pakull, 
Ivo Saviane, Curtis Saxton, Kinwah Wu for illuminating discussions. 
In particular, we are very grateful to Sean Farrell and Mat Servillat 
for sharing and discussing their {\it HST} results in advance of publications.
We thank the anonymous referee for a number of useful suggestions 
and constructive criticism.
RS acknowledges support from a Curtin University Senior Research Fellowship, 
and hospitality at the ESO headquarters in Santiago (Chile), 
at the Mullard Space Science Laboratory (UK) and 
at the University of Sydney (Australia) during part of this work. 
JCG acknowledges support from the Avadh Bhatia Fellowship and 
Alberta Inginuity.


\begin{thebibliography}{99}

\bibitem[\protect\citeauthoryear{Arnaud}{1996}]{arn96}
  Arnaud K.A., 1996, ASP Conference Series, Vol. 101, 
  Astronomical Data Analysis Software and Systems V., 
  Astron. Soc. Pac., San Francisco, G. Jacoby and J. Barnes, eds., p. 17

\bibitem[\protect\citeauthoryear{Bekki \& Freeman}{2003}]{bek03}
  Bekki K., Freeman K.C., 2003, MNRAS, 346, L11


\bibitem[\protect\citeauthoryear{Bertin}{2006}]{ber06}
  Bertin E., 2006, Automatic Astrometric and Photometric Calibration 
  with {\small{SCAMP}}, ASP Conference Series, Vol. 351, C. Gabriel, 
  C. Arviset, D. Ponz, and E. Solano, eds., p. 112

\bibitem[\protect\citeauthoryear{Bertin \& Arnout}{1996}]{ber96}
  Bertin E., Arnouts S., 1996, A\&AS, 317, 393

\bibitem[\protect\citeauthoryear{Bertin et al.}{2002}]{ber02}
  Bertin E., Mellier Y., Radovich M., Missonnier G., Didelon P., Morin, B.,
  2002, ASP Conference Series, Vol. 281, D.A. Bohlender, D. Durand, 
  and T.H. Handley, eds., p. 228

\bibitem[\protect\citeauthoryear{Bessell}{1990}]{bes90}
  Bessell M.S., 1990, PASP, 102, 1181

\bibitem[\protect\citeauthoryear{Davis et al.}{2011}]{dav11}
	Davis S.W., Narayan R., Zhu Y., Barret D., Farrell S.A., 
        Godet O., Servillat M., Webb N.A., 2011, ApJ, 734, 111

\bibitem[\protect\citeauthoryear{de Jong et al.}{1996}]{dej96}
  de Jong J.A., van Paradijs J., Augusteijn T., 1996, A\&A, 314, 484

\bibitem[\protect\citeauthoryear{Di Stefano et al.}{2003}]{dis03}
	Di Stefano R., Kong A.K.H., VanDalfsen M.L., Harris W.E., 
        Murray S.S., Delain K.M. 2003, ApJ, 599, 1067

\bibitem[\protect\citeauthoryear{Dubus et al.}{1999}]{dub99}
Dubus G., Lasota J.-P., Hameury J.-M., Charles P., 1999, MNRAS, 303, 139

\bibitem[\protect\citeauthoryear{Evans et al.}{2007}]{eva07}
  Evans P.A., et al., 2007, A\&A, 469, 379

\bibitem[\protect\citeauthoryear{Evans et al.}{2009}]{eva09}
  Evans P.A., et al., 2009, MNRAS, 397, 1177

\bibitem[\protect\citeauthoryear{Farrell et al.}{2009}]{far09} 
  Farrell S.A., Webb N.A., Barret D., Godet O., Rodrigues J.M., 2009, 
  Nature, 460, 73

\bibitem[\protect\citeauthoryear{Farrell et al.}{2011}]{far11} 
  Farrell S.A., et al., 2011, ApJ, submitted (arXiv:1110.6510)

\bibitem[\protect\citeauthoryear{Freitag et al.}{2006}]{fre06}
	Freitag M., G\"urkan M.A., Rasio, F.A., 2006, MNRAS, 368, 141

\bibitem[\protect\citeauthoryear{Frew}{2004}]{fre04}
  Frew D.J., 2004, JAD, 10, 6

\bibitem[\protect\citeauthoryear{Gebhardt et al.}{2005}]{geb05}
    Gebhardt K., Rich R.M., Ho L.C., 2005, ApJ, 634, 1093

\bibitem[\protect\citeauthoryear{Gierli\'nski et al.}{2008}]{gie08}
  Gierli\'nski M., Done C., Page K, 2008, MNRAS, 388, 753

\bibitem[\protect\citeauthoryear{Gierli\'nski et al.}{2009}]{gie09}
  Gierli\'nski M., Done C., Page K, 2008, MNRAS, 392, 1106

\bibitem[\protect\citeauthoryear{Godet et al.}{2009}]{god09}
	Godet O., Barret D., Webb N.A., Farrell S.A., Gehrels, N., 
        2009, ApJ, 705, L109

\bibitem[\protect\citeauthoryear{G\"uver \& \"Ozel}{2009}]{guv09}
    G\"uver T., \"Ozel F., 2009, MNRAS, 400, 2050

\bibitem[\protect\citeauthoryear{Hynes et al.}{2002}]{hyn02}
  Hynes R.I., Haswell C.A., Chaty S., Shrader C.R., Cui W. 2002, MNRAS, 331, 169

\bibitem[\protect\citeauthoryear{Hynes}{2005}]{hyn05}
  Hynes R., 2005, ApJ, 623, 1026

\bibitem[\protect\citeauthoryear{Kalberla et al.}{2005}]{kal05} 
  Kalberla P.M.W., Burton W.B., Hartmann D., Arnal E.M., Bajaja E., 
  Morras R., P\"oppel W.G.L.,  2005, A\&A, 440, 775

\bibitem[\protect\citeauthoryear{King \& Dehnen}{2005}]{kin05}
  King A.R., Dehnen W., 2005, MNRAS, 357, 275

\bibitem[\protect\citeauthoryear{King \& Ritter}{1998}]{kin98}
  King A.R., Ritter H., 1998, MNRAS, 293, L42

\bibitem[\protect\citeauthoryear{Lasota et al.}{2011}]{las11}	
	Lasota J.-P., Alexander T., Dubus G., Barret D., Farrell S.A., 
        Gehrels N., Godet O., Webb N.A., 2011, ApJ, 735, 89

\bibitem[\protect\citeauthoryear{Leitherer et al.}{1999}]{lei99}
    Leitherer C., et al., 1999, ApJS, 123, 3

\bibitem[\protect\citeauthoryear{L\"utzgendorf et al.}{2011}]{lut11}
    L\"utzgendorf N., Kissler-Patig M., Noyola E., Jalali B., de Zeeuw P.T., 
    Gebhardt K., Baumgardt H., 2011, A\&A, 533, 36

\bibitem[\protect\citeauthoryear{Massey et al.}{2001}]{mas01}
	Massey P., DeGioia-Eastwood K., Waterhouse E., 2001, AJ, 121, 1050

\bibitem[\protect\citeauthoryear{Nonino et al.}{2009}]{non09} 
  Nonino M., et al., 2009, ApJSS, 183, 244

\bibitem[\protect\citeauthoryear{O'Leary \& Loeb}{2009}]{ole09}
  O'Leary R.M., Loeb A., 2009, MNRAS, 395, 781O

\bibitem[\protect\citeauthoryear{Peng et al.}{2002}]{pen02}
  Peng C.Y., Ho L.C., Impey C.D., Rix H.-W., 2002, AJ, 124, 266

\bibitem[\protect\citeauthoryear{Peng et al.}{2010}]{pen10}
  Peng C.Y., Ho L.C., Impey C.D., Rix H.-W., 2010, AJ, 139, 2097

\bibitem[\protect\citeauthoryear{Pickles}{1998}]{pic98}
  Pickles A.J., 1998, PASP, 110, 863

\bibitem[\protect\citeauthoryear{Pizzolato et al.}{2010}]{piz10}
	Pizzolato F., Wolter A., Trinchieri G.,
        2010, MNRAS, 406, 1116

\bibitem[\protect\citeauthoryear{Portegies Zwart \& McMillan}{2002}]{por02}
	Portegies Zwart S.F., McMillan S.L.W., 2002, ApJ, 576, 899

\bibitem[\protect\citeauthoryear{Schlegel et al.}{1998}]{sch98}
	Schlegel D.J., Finkbeiner D.P., Davis M.,
        1998, ApJ, 500, 525


\bibitem[\protect\citeauthoryear{Servillat et al.}{2011}]{ser11}
  Servillat M., Farrell S.A., Lin D., Godet O., Barret D., Webb N., 
  2011, ApJ, to appear in the 2011 December 10 issue (arXiv:1108.4405)

\bibitem[\protect\citeauthoryear{Shakura \& Sunyaev}{1973}]{sha73}
  Shakura N.I., Sunyaev R.A., 1973, A\&A, 24, 337

\bibitem[\protect\citeauthoryear{Soria et al.}{2010}]{sor10}
	Soria R., Hau G.K.T., Graham A.W., Kong A.K.H., Kuin N.P.M., 
        Li I.-H., Liu J.-F., Wu K., 2010, MNRAS, 405, 870

\bibitem[\protect\citeauthoryear{Swartz et al.}{2004}]{swa04}
  Swartz D.A., Ghosh K.K., Tennant A.F., Wu K., 2004, ApJS, 154, 519

\bibitem[\protect\citeauthoryear{Swartz et al.}{2011}]{swa11}
  Swartz D.A., Soria R., Tennant A.F., Yukita M., 
  2011, ApJ, 741, 49

\bibitem[\protect\citeauthoryear{van der Marel \& Anderson}{2010}]{van10} 
  van der Marel R.P., Anderson J., 2010, ApJ, 710, 1063

\bibitem[\protect\citeauthoryear{V\'azquez \& Leitherer}{2005}]{vaz05}
	V\'azquez G.A., Leitherer C., 2005, ApJ, 621, 695

\bibitem[\protect\citeauthoryear{Vrtilek et al.}{1990}]{vrt90} 
	Vrtilek S.D., Raymond J.C., Garcia M.R., Verbunt F., Hasinger G., 
        Kurster M. 1990, A\&A, 235, 162

\bibitem[\protect\citeauthoryear{Walton et al.}{2011}]{wal11}
	Walton D.J., Roberts T.P., Mateos S., Heard V.,
        2011, MNRAS, 416, 1844

\bibitem[\protect\citeauthoryear{Wiersema et al.}{2010}]{wie10}
  Wiersema K., Farrell S.A., Webb N.A., Servillat M., Maccarone T.J., 
  Barret D., Godet O., 2010, ApJ, 721, L102

\bibitem[\protect\citeauthoryear{Worthey}{1994}]{wor94}
  Worthey G., 1994, ApJS, 95, 107

\end{thebibliography}
\end{document}